\documentclass[%
  twocolumn,%
  groupedaddress,%
  showpacs,%
  lengthcheck,%
  letterpaper,%
  floatfix%
]{revtex4}

\usepackage{hyperref,amsmath,amssymb,graphicx,bm}
\usepackage{fix-cm,type1cm} 

\hypersetup{%
  breaklinks = {true},
  citecolor = {blue},
  colorlinks = {true},
  linkcolor = {red},
  pdfauthor = {\textcopyright\ Vitor M. Pereira},
  pdfcreator = {\LaTeX\ and \flqq hyperref\frqq},
  pdffitwindow = {true},
  pdfmenubar = {true},
  pdfpagelayout = {SinglePage},
  pdfstartview = {Fit},
  pdftoolbar = {true},
  plainpages = {false},
}

\newcommand{\vF}{\ensuremath{\nu_\text{F\,}}}
\newcommand{\bsigma}{\ensuremath{\bm\sigma}}
\newcommand{\bp}{\ensuremath{\bm {p}}}
\newcommand{\br}{\ensuremath{\bm {r}}}
\newcommand{\Hcal}{\mathcal{H}}
\newcommand{\Mcal}{\mathcal{M}}
\newcommand{\Ncal}{\mathcal{N}}
\newcommand{\gc}{g_\text{c}}
\newcommand{\e}{\epsilon}
\newcommand{\gt}{\tilde{g}}

\DeclareMathOperator{\sign}{sign}
\DeclareMathOperator{\real}{Re}

\begin{document}


\title{The Coulomb impurity problem in graphene}


\author{Vitor M. Pereira}
\affiliation{Department of Physics, Boston University, 590 
Commonwealth Avenue, Boston, MA 02215, USA}

\author{Johan Nilsson}
\affiliation{Department of Physics, Boston University, 590 
Commonwealth Avenue, Boston, MA 02215, USA}

\author{A.~H. Castro Neto}
\affiliation{Department of Physics, Boston University, 590 
Commonwealth Avenue, Boston, MA 02215, USA}

\date{\today}


\begin{abstract}
We address the problem of an unscreened Coulomb charge in graphene, and
calculate the local density of states and displaced charge as a
function of energy and distance from the impurity. This is done
non-perturbatively in two different ways: (1) solving the problem exactly by
studying numerically the tight-binding model on the lattice;
(2) using the continuum description in terms of the 2D Dirac
equation. We show that the Dirac equation, when properly regularized,
provides a qualitative and quantitative low energy description of the problem. 
The lattice solution shows extra features that cannot
be described by the Dirac equation, namely bound state formation and
strong renormalization of the van~Hove singularities.
\end{abstract}

\maketitle



Since the isolation of graphene a few years ago
\cite{Geim_review} there is an intense interest in understanding the
electronic properties of this material. The low energy 
electronic excitations in graphene are linearly dispersing, massless,
chiral Dirac fermions, described by Dirac cones at the edges of the
Brillouin zone (BZ) (at the K and K$^\prime$ points). 
Due to the vanishing density of states of these Dirac fermions in undoped
graphene, the system is very sensitive to impurities and defects \cite{nuno},
which control its transport properties \cite{nomura,shafique}. This also means
that electrons in graphene screen poorly, and hence the issue of
unscreened Coulomb interactions becomes paramount in the understanding of
the experimental data \cite{Geim_review}.

In this paper we contrast the tight-binding approach (that we solve
exactly with numerical techniques) with the continuum approach 
based on the Dirac equation. We show that the latter
provides a good qualitative description of the problem at low
energies, when properly regularized. We also show
that the Dirac description fails at moderate to high energies and at
short distances, when the lattice description is the only one possible.
In this case new features, not captured by the Dirac
Hamiltonian, emerge. We calculate the local density of
states (LDOS) and induced charge around a Coulomb impurity as a function
of energy and distance. These quantities are experimentally accessible through
scanning tunneling spectroscopy (STS). We stress that calculations for
impurities with long-range potentials are radically
different from the ones for short-range forces, which are exactly solvable
using T-matrix methods \cite{nuno_boron}.

Consider the problem of a single Coulomb impurity, with charge $Z e$, 
placed in the middle of an hexagon of the honeycomb lattice \cite{Note:4}.
The tight-binding Hamiltonian for this problem, with nearest-neighbor hopping
only, is given by (we use units such that $\hbar =1$): 
\begin{eqnarray}
  \Hcal = t \sum_{i} 
    \bigl(a^\dagger_i b_i + \text{h.c.}\bigr) + 
    \frac{Z e^2}{\varepsilon_0} \sum_{i} 
      \biggl(
        \dfrac{a^\dagger_i a_i}{r^A_i} + \dfrac{b^\dagger_i b_i}{r^B_i}
      \biggr)
   ,
  \label{eq:H-Lattice} 
\end{eqnarray}
where $a_i$ ($b_i$) annihilates an electron at site ${\bf R}_i$ and sublattice
$A$ ($B$), $t \approx 2.7$~eV is the hopping energy,
and $r_i^{A,B}$ is the distance between the carbon atoms and the impurity
(assumed to be at the origin of the coordinate system); $\varepsilon_0$ is the
dielectric constant. 
We calculated numerically the spectrum of \eqref{eq:H-Lattice} using 
the methods of exact diagonalization and recursion used in 
ref.~\cite{Pereira:2006} for the study of short-range unitary scatterers.

Close to the K point in the BZ we can write an effective low energy
Hamiltonian for \eqref{eq:H-Lattice} in terms of Dirac fermions with a spinor
wave function $\Psi({\br})$, whose components represent its weight on each
sublattice. The wave function obeys the equation:
\begin{equation}
  \vF \Bigl(
    \bsigma\cdot \bp - g/r  
  \Bigr)
  \Psi(\br) = E\;\Psi(\br)
  \label{eq:SchrodingerEq}
  ,
\end{equation}
with $\vF = 3at/2$ ($\approx 10^6$ m/s) being the Fermi
velocity, \bp\ the 2D momentum operator, $\sigma_i$ the Pauli matrices,
and $g \equiv Z e^2/(\vF\varepsilon_0)$ is the dimensionless coupling
constant. Henceforth, we shall 
take $a$ (the C--C distance) and \vF as distance and energy units. Notice that
\eqref{eq:SchrodingerEq} does not involve
inter-cone scattering because the unscreened Coulomb potential is dominated by 
small momentum transfers since, in Fourier space, it behaves like $1/q$ and is
singular as $q \to 0$. 

Eq.~\eqref{eq:SchrodingerEq} is separable in cylindrical coordinates. Resorting
to eigenfunctions of the conserved angular momentum, 
$J_z=L_z + \sigma_z /2$ \cite{DiVincenzo:1984},
\begin{equation}
  \psi_j(\br) = \frac{1}{\sqrt{r}} 
  \left(
  \begin{array}{c}
    e^{i(j-\frac{1}{2})\varphi} \, \varphi^A_j(r) \\    
    i e^{i(j+\frac{1}{2})\varphi} \, \varphi^B_j(r)  
  \end{array}
  \right)
  , 
  \label{eq:Def-Spinor}
\end{equation}
the radial equation for \eqref{eq:SchrodingerEq} reads
$(j=\pm1/2,\,\pm3/2,\,\ldots)$

\begin{equation}
\left[
\begin{array}{cc}
    \e\!+\!g/r \!&\! -(\partial_r\!+\!j/r) \\
    (\partial_r\!-\!j/r) \!&\! \e\!+\!g/r
\end{array}  
\right]
\left[
\begin{array}{c}
  \varphi^A_j \\
  \varphi^B_j
\end{array}  
\right]
  \equiv
  \Mcal_j \varphi_j(r)
  =0
  .
  \label{eq:SchrodingerEq-Radial}
\end{equation}
This equation can be solved by multiplication on the left by
$\Mcal_j'=\sigma_z\Mcal_j\sigma_z$ and subsequent diagonalization. The
eigenstates are
then linear combinations of the type
\begin{equation}
  \varphi_j(r) = \sum_{\lambda =\pm} C_\lambda u_\lambda f_\lambda(r) 
  \,,\,
  u_\pm = \sqrt{\frac{1}{2|j|}} 
    \binom{\sqrt{|j\pm\alpha|}}{s_{gj} \sqrt{|j\mp\alpha|}}
  ,
  \label{eq:DiagonalWF}
\end{equation}
where $s_x \equiv \sign(x)$, $\alpha=\sqrt{j^2-g^2}$ and $f_\lambda(r)$ solves
\begin{equation}
  \partial_r^2 f_\lambda(r) + 
  \left[
    \e^2 + 2g\e/r - \alpha(\alpha-\lambda)/r^2
  \right]
  f_\lambda(r) = 0
  .
  \label{eq:CoulombEq}
\end{equation}
Introducing $\rho=|\e|r$, the above becomes the familiar radial equation for the
3D Coulomb problem \cite{Landau-QM:1981}, and the presence of $\e^2$ (rather
than $\e$) entails the absence of bound solutions in the Dirac problem. 
It is important to note that
when $g$ is above $\gc=1/2$, the parameter $\alpha$ in eq.~\eqref{eq:CoulombEq}
becomes imaginary for some angular momentum channels. The nature of the
solutions is then radically different, which, as will be seen, has dramatic
consequences. We address the two regimes separately.


When $g<\gc$, eq.~\eqref{eq:CoulombEq} can be solved in terms of Coulomb
wave functions \cite{Abramowitz:1964,Yost:1936}: $F_L(\eta,\rho)$,
$G_L(\eta,\rho)$. In fact, letting $\gt=s_\e g$, it is straightforward to show
that the appropriate linear combination in \eqref{eq:DiagonalWF} that solves
\eqref{eq:SchrodingerEq-Radial} is
\begin{equation}
  \varphi_j(r) / \Ncal_j = 
    u_+ F_{\alpha-1}(-\gt,\rho) + s_{g\e} u_- F_{\alpha}(-\gt,\rho)
  ,
  \label{eq:RadialSolution}
\end{equation}
where only the regular solution at the origin has been included. Since
$F_\alpha(-\gt,\rho)$ are the regular scattering solutions of the 3D Coulomb
problem, they include the well-known logarithmic phase shift in the
asymptotic expansion
\cite{Landau-QM:1981}:
\begin{eqnarray}
  F_\alpha(-\gt,\rho) \sim \sin\Bigl(
    \rho + \gt \log(2\rho) + \vartheta_\alpha(\gt)
  \Bigr), 
  \label{eq:Coulomb-Phase-Shift}  
\end{eqnarray}
where $\vartheta_\alpha(\gt) = - \alpha \frac{\pi}{2} +
    \arg\bigl[\Gamma(1+\alpha-i\gt)\bigr]$.
The logarithmic phase shift also carries to our case, for
\eqref{eq:RadialSolution} can always be written asymptotically as 
\begin{equation}
  \sin\Biggl[ 
    \rho + \gt \log(2\rho) +
    \arg\Bigl(
      u_+ e^{i\vartheta_{\alpha-1}}+s_{g\e} u_- e^{i\vartheta_{\alpha}}
    \Bigr)
  \Biggr]
  .
  \label{eq:Our-Phase-Shift}  
\end{equation}
The normalization, $\Ncal_j$, is determined by imposing orthogonality on
the energy scale, $\int
\psi_i(\e,\br)^\dagger\psi_j(\e',\br) d\br = \delta_{ij}\delta(\e-\e')$, leading
to $\Ncal_j^{-2}=2\pi^2\alpha^2/j^2$. With
this choice, one conveniently recovers the free DOS per unit area and cone when 
$\gt=0$. To see this, one notes that the LDOS, $N(\epsilon,r) =
\sum_E |\Psi_E(r)|^2 \delta(\epsilon-E)$, is given by
$
  N(\e,\br) = \sum_{j=-\infty}^{\infty} n_j(\e,r).
$
Using $n_j(\e,r) \equiv r^{-1} |\varphi_j^A(r)|^2 + r^{-1}
|\varphi_j^B(r)|^2$, the contribution from each angular momentum channel is
simply:
\begin{equation}
  n_j(\e,r)\!=\!(\Ncal_j^2/r)
  \left[
    F_{\alpha-1}^2 + F_{\alpha}^2 + 2\gt F_{\alpha}F_{\alpha-1}/|j|
  \right]
  .
  \label{eq:LDOS-Weak}
\end{equation}
In the limit $\gt\to0$ the Coulomb wave functions reduce to Bessel functions
\cite{Abramowitz:1964}, and one obtains $N(\e,\br) = |\e|/2\pi$.

Of the several aspects encoded in \eqref{eq:LDOS-Weak}, two are immediate:
particle-hole symmetry is lost, and the LDOS becomes singular as $E\to 0$.
This last point follows from the fact that, in this limit, $N(\e,\br)\propto
|\e|^{2\alpha}$ and $\alpha<1/2$ for $|j|=1/2$; the asymmetry
stems from the dependence of $\gt$ on the sign of the energy.

\begin{figure}[t]
\begin{center}
\includegraphics*[width=1.0\columnwidth]{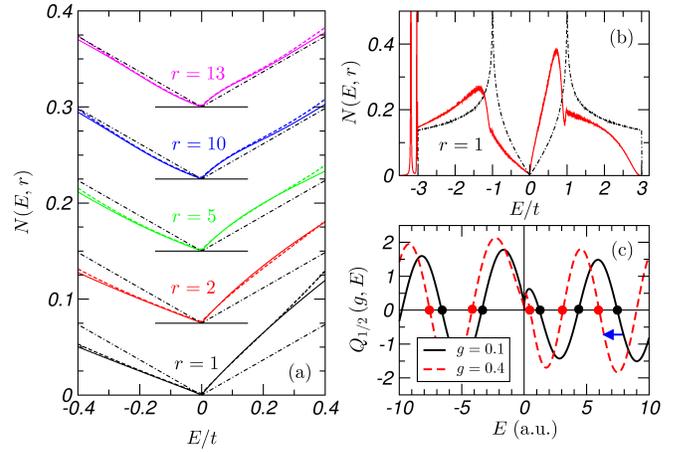}
\end{center}
\caption{(color online) 
  (a) Comparison of the LDOS (solid) with
      the numerical results in the lattice (dashed), calculated at different
      distances from the impurity and $g=1/6$. The DOS for $g=0$ is also
      included for comparison (dot-dashed). For clarity, curves at different $r$
      have been vertically displaced. 
  (b) LDOS at the site closest to the impurity in the lattice for $g=1/3$.
  (c) Quantization condition \eqref{eq:Quantization} with $j=1/2$, for
      $g=0.1\text{ and }0.4$.
  }
\label{fig:Fig1}
\end{figure}

It is most instructive to compare the results derived within the Dirac
approximation \eqref{eq:SchrodingerEq}, with the results on the
lattice that one obtains using the full Hamiltonian in
eq.~\eqref{eq:H-Lattice}. 
In Fig.~\ref{fig:Fig1}(a) one can observe that, at low
energies ($E\lesssim 0.5 t$), the result of \eqref{eq:LDOS-Weak} reproduces the
LDOS on the lattice even at distances of the order of the lattice
parameter (the two cases are barely distinguishable for most of the plotted
range). Moreover, the attractive Coulomb potential brings locally a reduction of
spectral weight in the lower band, the opposite happening to the upper band. The
effect is strongest near the impurity and evolves towards the bulk behavior at
larger distances.

%
%

This behavior of the spectrum near the Dirac point can be understood from an
investigation of the quantized energies when the system is restricted to a
region of finite radius, $R$. 
A convenient way to confine 2D Dirac fermions is to
introduce an infinite mass at the boundary \cite{Berry:1987}, which translates
into the Boundary Condition (BC)
$
  \varphi_j^A(R)=\varphi_j^B(R) \,.
$
From \eqref{eq:RadialSolution} this expands into
\begin{equation}
  Q_j \equiv F_{\alpha-1}(-\gt,R|\e|) - s_j s_\e F_{\alpha}(-\gt,R|\e|) = 0
  .
  \label{eq:Quantization}
\end{equation}
This equation, whose roots determine the quantized energy levels, 
always has the 
trivial solution, $\e = 0$. As one can easily verify
[see Fig.~\ref{fig:Fig1}(c) and also the asymptotic phase shift in
\eqref{eq:Our-Phase-Shift}], the other nodes are simply shifted to lower
energies with increasing $g$ --- just as expected under an attractive potential
--- but they never cross $\e=0$. This means that no states are
phase-shifted 
across the Dirac point, but they rather heap up close to it when $\e>0$, and
conversely for $\e<0$. Hence, even though the gap in graphene is zero,
no states 
will cross it while $g<\gc$, much like in a conventional semiconductor. In the
spirit of Friedel's argument \cite{Friedel:1952}, this effect has profound
consequences for the induced charge, which will be discussed later. Finally, we
remark that, although the continuum approximation does not support bound
solutions (unless a cutoff is introduced), they appear naturally in the lattice.
This is seen in Fig.~\ref{fig:Fig1}(b), where a bound state barely detached from
the band is signaled by the sharp peak at the lower band edge. Furthermore,
notice how the van Hove singularities are strongly renormalized by the presence
of the impurity.


For $g>g_c$, $\alpha$ can become purely imaginary, and we
introduce $\beta=-i\alpha = \sqrt{g^2-j^2}$ for those $j$'s such that
$|g|>|j|$. In general, linearly independent solutions of \eqref{eq:CoulombEq}
are \cite{Note:1}
\begin{align}
  \lambda = +1:&\quad F_{\alpha-1}(-\gt,\rho)~\text{ and }
F_{-\alpha}(-\gt,\rho), \\
  \lambda = -1:&\quad F_{\alpha}(-\gt,\rho)~\text{ and }
F_{-\alpha-1}(-\gt,\rho)
  .
  \label{eq:CoulombSolutionsInGeneral}
\end{align}
When $\alpha\in\mathbb{R}$ the ones with negative index are divergent at the
origin and thus only the first were kept in \eqref{eq:RadialSolution}. But when
$\alpha\in i\mathbb{R}$, the solutions are well behaved at the origin (albeit
oscillatory), and two linearly independent solutions emerge. One is analogous
to \eqref{eq:RadialSolution}:
\begin{equation}
  \bar{\varphi}_{i\beta}(r)  = 
    \bar{u}_+ F_{i\beta-1}(-\gt,\rho) + 
    s_{jg\e} \bar{u}_- F_{i\beta}(-\gt,\rho)
  ,
  \label{eq:RadialSolution-2}
\end{equation}
apart from a normalization factor, and where now
\begin{equation}
  \bar{u}_\pm = \sqrt{\frac{1}{2|g|}} 
  \binom{\sqrt{j\pm i\beta}}{s_{g} \sqrt{j\mp i\beta}}
  .
  \label{eq:DiagonalWF-2}
\end{equation}
The other solution is simply $\bar{\varphi}_{-i\beta}(r)$.
The general solution is therefore of the type
$
  \bar{\varphi}_j(r) = C_1 \; \bar{\varphi}_{i\beta}(r) + 
    C_2 \; \bar{\varphi}_{-i\beta}(r) \, ,
$
where $C_{1,2}$ are to be set by the BC at short distances.
Since we seek the effective low energy description of a problem defined in a
lattice, a natural BC
is to have an infinite mass at some short cutoff distance $a_0 \simeq a$. This
has 
the effect of forbidding the penetration of electrons to distances shorter than
$a_0$ \cite{Berry:1987}, thus reflecting the physical situation,
while, at the same time, naturally curing the divergence in the
potential at the 
origin. This translates again into a BC
$\varphi_j^A(a_0)=\varphi_j^B(a_0)$,
and given that $C_{1,2}$ can always be chosen so 
that $C_1/C_2=\exp[2i\delta_j(\e)]$, one then obtains the phase $\delta_j(\e)$:
\begin{equation}
  e^{i 2 \delta_j(\e)} = s_g 
    \frac{
      F_{-i\beta-1} - s_{\e j} F_{-i\beta }
    }{
      F_{i\beta-1} - s_{\e j} F_{i\beta}
    } \Bigl|_{\rho = \e a_0}
  .
  \label{eq:PhaseShift}  
\end{equation}
We can follow the same procedure as before to normalize the states
in the energy scale, and then extract the contribution of the 
overcritical $j$'s to the LDOS:
\begin{equation}
  \bar{n}_j(\e,r) = \frac{1}{2 \pi^2 r}
  \frac{
    \varrho_j^I(\rho) + 
    s_{\e j} \real\bigl[e^{i 2 \delta_j } \varrho_j^{II}(\rho) \bigr]
  }{
    \Bigl\langle
    \varrho_j^I(\infty) + 
    s_{\e j} \real\bigl[e^{i 2 \delta_j } \varrho_j^{II}(\infty) \bigr]
    \Bigr\rangle_r
  }
  ,
  \label{eq:LDOS-Strong}  
\end{equation}
where, for readability, we defined  
\begin{align*}
    \varrho_j^I &\equiv
    |F_{i\beta }|^2  + |F_{i\beta -1}|^2 + 
    \frac{2|j|}{\gt} \real[F_{i\beta} F_{-i\beta -1}]
    ,\\
    \varrho_j^{II} &\equiv 
    2 F_{i\beta} F_{i\beta -1} + \frac{|j|}{\gt}
    (F_{i\beta }^2 + F_{i\beta -1}^2)
    ,
\end{align*}
and $\langle\cdots\rangle_r$ stands for the constant term as $r\to\infty$.
Eqs.~\eqref{eq:LDOS-Weak} and \eqref{eq:LDOS-Strong} determine the LDOS for any
coupling strength, $g$, which can be summarized as 
\begin{equation}
  N(\e, \br) = \sum_{|j|<|g|} \bar{n}_j(\e,r) + \sum_{|j|>|g|} n_j(\e,r)
  .
  \label{eq:LDOS-Any}
\end{equation}

The presence of the first term in eq.~\eqref{eq:LDOS-Any} brings a profound
rearrangement of the spectrum close to the impurity, with much more striking
consequences than in the weak coupling regime.
%
\begin{figure}[t]
\begin{center}
\includegraphics*[clip,width=1.0\columnwidth]{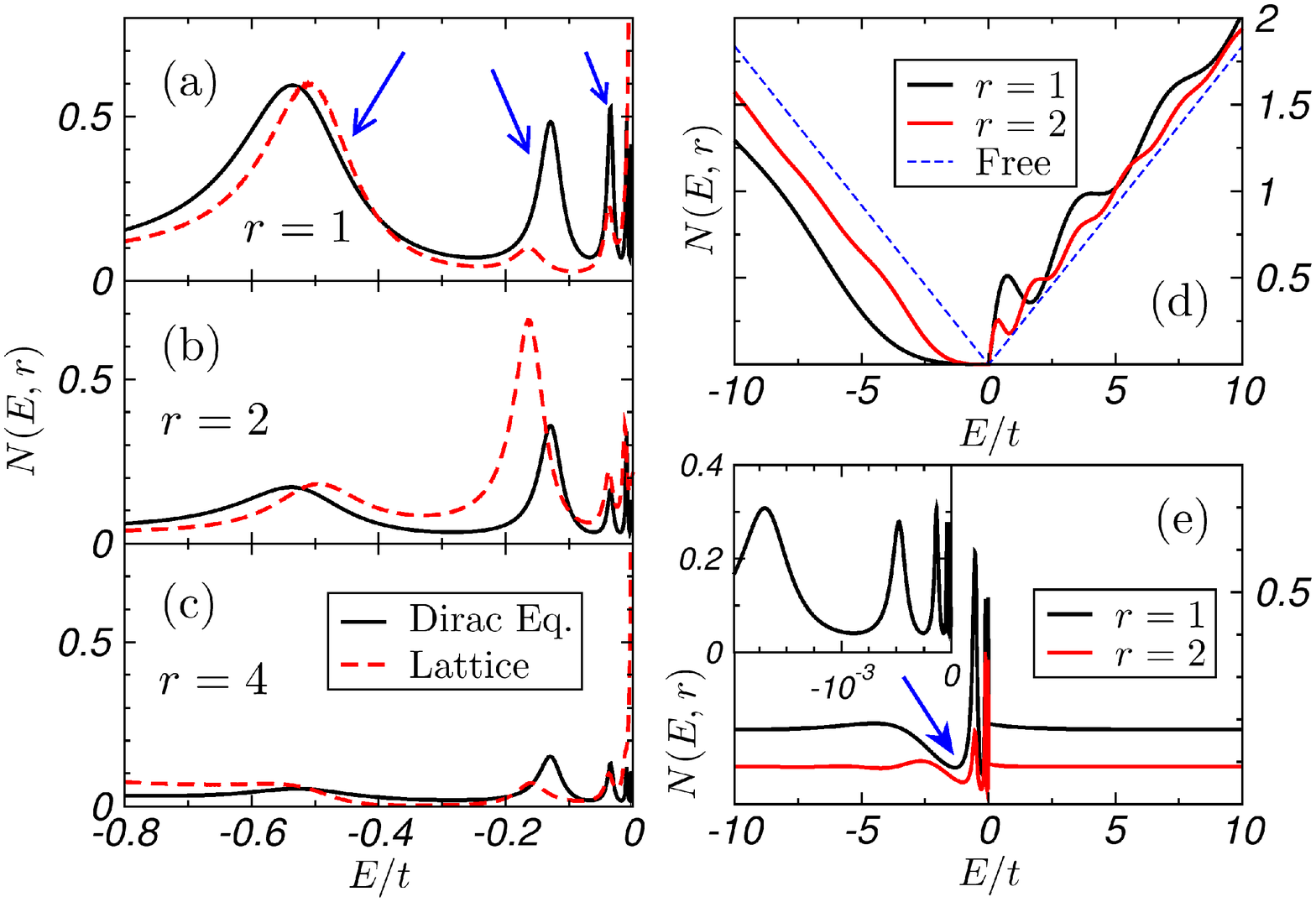}
\end{center}
\caption{(color online)
  (a--c) The LDOS in the lattice (dashed, recursion method) is compared with the
        first contribution in \eqref{eq:LDOS-Any} (solid) for 
        different distances from the impurity.
  (d) The second contribution in \eqref{eq:LDOS-Any} (solid) and
      the free, linear, DOS for reference.
  (e) First contribution in \eqref{eq:LDOS-Any}; the inset is a 
      magnification for $\e\simeq 0$.
      In all panels $g=4/3$.
  }
\label{fig:Fig2}
\end{figure}
%
%
In Fig.~\ref{fig:Fig2}(a--c) we plot the LDOS obtained numerically in the
lattice with $g=4/3$, together with the first contribution in
\eqref{eq:LDOS-Any}. For such $g$ it comes only from $\bar{n}_{\pm1/2}(\e,r)$,
and we used $a_0=0.55 a$ to impose the BC. It is clear that the analytical
result captures quite accurately the behavior of the LDOS in the lattice.
Most importantly, both results exhibit 3 \emph{marked resonances} in the
negative (hole-like) energy region, which decay away from the impurity. Indeed,
their amplitude is such that they dominate the profile of the LDOS at low
energies. Increasing $g$ will cause the resonances to migrate
downwards in energy, and their number to increase. This is rather peculiar and
has to do
with the fact that, in reality, the Dirac point is an accumulation point of
infinitely many resonances [inset in Fig.~\ref{fig:Fig2}(e)]. One can
appreciate 
the origin of this from the fact that $F_{L-1}(\eta,\rho\simeq0) \sim
\rho^{L}$.
Since $L\in \mathbb{C}$ in eqs.~\eqref{eq:RadialSolution-2}
and~\eqref{eq:PhaseShift}, it implies that the
wave functions oscillate with logarithmically diverging frequency as
$\e\to0$. This situation is akin to the \emph{fall of a particle to the center}
\cite{Landau-QM:1981}, and the effect carries to the LDOS with the consequences
shown in Figs.~\ref{fig:Fig2}(a--c). In panel (d), we present the remainder
contribution (second term) to the total LDOS in eq.~\eqref{eq:LDOS-Any}. It is
evident that in the region $\e\lesssim0$, dominated by the resonances, this
contribution is highly suppressed, whereas, for positive energies, the LDOS 
exhibits an oscillating behavior around the bulk limit. 

%
\begin{figure}[t]
\begin{center}
\includegraphics*[clip,width=1.0\columnwidth]{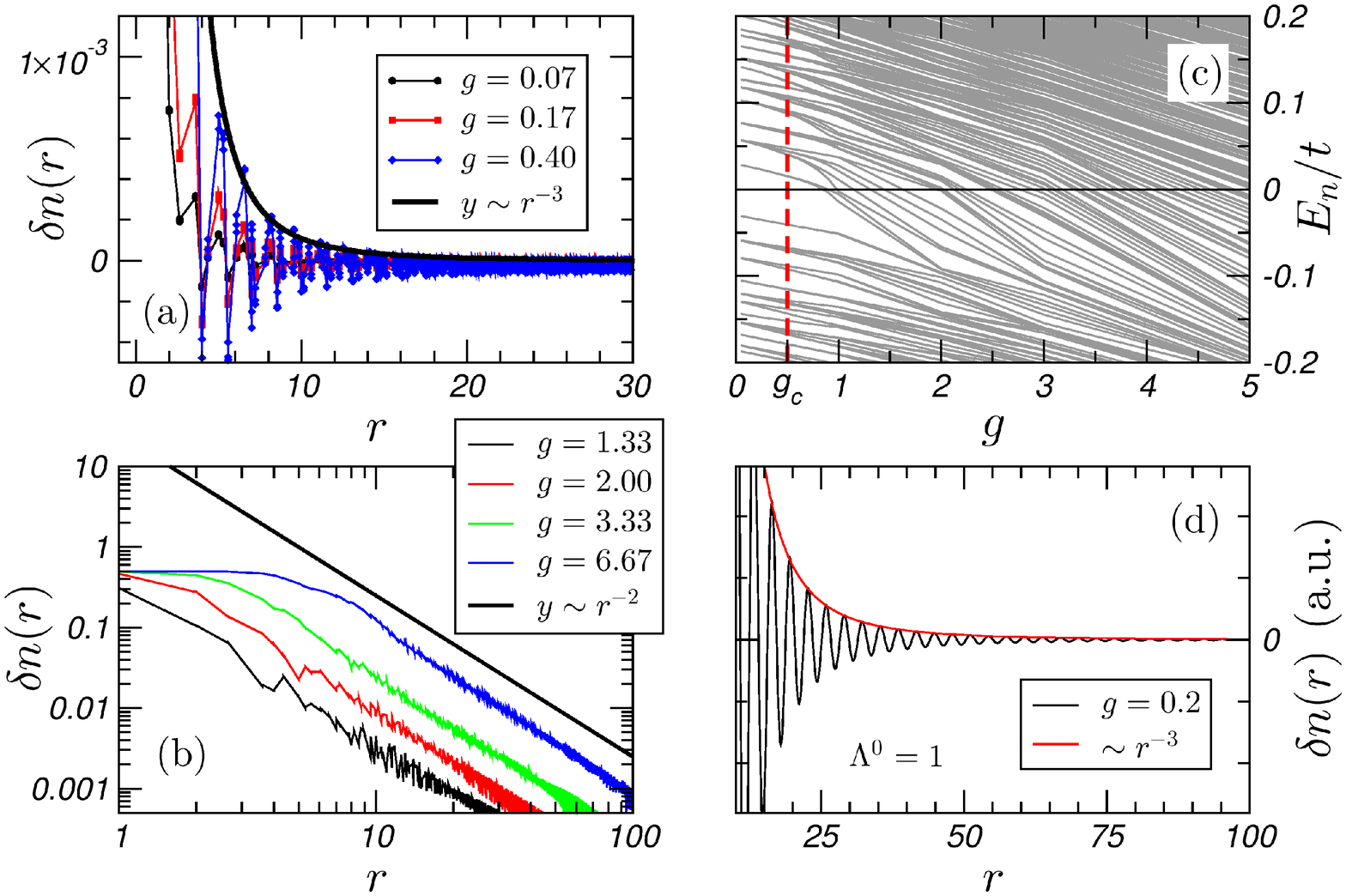}
\end{center}
\caption{(color online)
  (a) Induced charge numerically obtained from exact diagonalization on a
  lattice with $124^2$ sites ($g<\gc$).
  (b)  \emph{Idem} ($g>\gc$).
  (c) Evolution of the numerical spectrum with $g$.
  (d) Analytical $\delta n(r)$ obtained using \eqref{eq:LDOS-Integral} and
      regularization.
  }
\label{fig:Fig3}
\end{figure}
%
%


As is customary, it is also of interest here to understand
how the electronic density readjusts itself in the presence of this charged
impurity. Even though interactions are not included (and thus there is
no real screening), one can obtain important insights from the non-interacting
 problem in the spirit of Friedel \cite{Friedel:1952}. The charge density,
$n(r)$, is straightforwardly obtained from integration of
\eqref{eq:LDOS-Any} in energy from an energy cut-off, 
$-\Lambda$, up to $E_F=0$. Since it involves integrals of
$F_\alpha(-\gt,\rho)$, this can be done exactly. For example, when $g<\gc$,
one has (for each $j$ channel)
\begin{equation}
  n_j(r) =  
  \left[ \rho n_j(\e,r)/r - 
  |j| F_{\alpha} F_{\alpha-1}/(\pi^2 r^2) \right|_{\e=-\Lambda}
  .
  \label{eq:LDOS-Integral}
\end{equation}
Although the above needs only to be evaluated at $\e=-\Lambda$, it is not free
from difficulties yet, for there is an infinite sum over $j$ to be performed.
Expanding \eqref{eq:LDOS-Integral} asymptotically, the induced charge per
channel reads
\begin{equation}
  \delta n_j(r) \!=\! n_j(r) \!-\! n_j^0(r) \!\sim\! (1/r)
    \left[
      \Lambda \!-\! g/r \!-\! \Lambda^0 \!+\! \mathcal{O}(r^{-2})
    \right]
  ,
  \label{eq:InducedCharge-Weak}
\end{equation}
where the remainder is oscillating with frequency $\Lambda$, and convergent
with $j$. Clearly, if $\Lambda = \Lambda^0$ the sum over $j$ diverges. 
We regularize this by locally changing the cutoff:
$\Lambda = \Lambda^0 + g/r$, whereupon the leading contribution is 
$\sim r^{-3}$, and oscillating with frequency $\Lambda$
[Fig.~\ref{fig:Fig3}(d)]. Nonetheless, despite accidentally reproducing the
lattice behavior, the oscillation itself is tied to the cutoff procedure. We
point out
that any charge oscillation decaying faster than $1/r^2$ on the lattice appears,
in the continuum theory, as a Dirac delta function at the origin, in 
agreement with perturbative studies of this problem \cite{Kolezhuk:2006}, but
differs from the self-consistent calculation in ref.~\cite{DiVincenzo:1984}.
Interestingly, the behavior of the induced charge in the lattice is indeed
$\sim r^{-3}$ and oscillating, as seen in Fig.~\ref{fig:Fig3}(a), wherein exact
numerical results in the lattice are plotted. 
An analogous analytical procedure can be undertaken for $g>\gc$, leading to an
induced charge decaying as $\sim r^{-2}$. Fig.~\ref{fig:Fig3}(b)
shows that this agrees with the numerical data in the lattice, where
$\delta n(r)\sim r^{-2}$, and non-oscillating. One
thus concludes that the induced charge behaves \emph{quite differently} below
and above $\gc$, as had been hinted before on account of the peculiar behavior
of the phase shifts below $\gc$. This last point can be confirmed by inspecting
the behavior of the numerical energy levels as a function of $g$ shown in
Fig.~\ref{fig:Fig3}(c), being evident the difference between the two regimes.


We have studied the problem of a Coulomb charge in graphene via exact numerical
methods on the lattice and  the Dirac Hamiltonian. We calculated the LDOS 
and local charge as a function of energy and distance from the impurity,
having found that the Dirac equation provides a qualitative description of the
problem at low energies. We found new features in the lattice description that
are beyond the Dirac equation: bound states and strong renormalization of the
van Hove singularities. We have also shown the existence of a critical coupling
$g_c$ separating the weak and strong coupling regimes, with radical differences
in the features of the LDOS. These results can be tested experimentally  through
STS measurements.


We acknowledge useful discussions with V. Kotov, S. Sachdev, and B. Uchoa.
V.M.P. is supported by FCT via SFRH/BPD/27182/2006 and POCI 2010 via
PTDC/FIS/64404/2006; and acknowledges the use of computational facilities
at CFP. A.H.C.N. was supported through NSF grant DMR-0343790.
\emph{Note added}: While preparing the manuscript, we became aware of two
preprints \cite{Shytov:2007,Novikov:2007} with a similar approach to this
problem.


\bibliography{graphene_coulomb}

\begin{thebibliography}{17}
\bibitem[{\citenamefont{{This geometry is chosen to preserve sublattice
  symmetry.}}()}]{Note:4}
\bibinfo{author}{\bibnamefont{{This geometry is chosen to preserve sublattice
  symmetry.}}}
\bibitem[{\citenamefont{{If $2L\notin\mathbb{Z}$, one can express
  $G_L(\eta,\rho)$ in terms of $F_L(\eta,\rho)$ and $F_{-L-1}(\eta,\rho)$, much
  like as in the Bessel functions.}}()}]{Note:1}
\bibinfo{author}{\bibnamefont{{If $2L\notin\mathbb{Z}$, one can express
  $G_L(\eta,\rho)$ in terms of $F_L(\eta,\rho)$ and $F_{-L-1}(\eta,\rho)$, much
  like as in the Bessel functions.}}}
\end{thebibliography}

\end{document}